\documentclass[conference]{IEEEtran}

\ifCLASSINFOpdf
\else
\fi
\usepackage{amsmath}
\usepackage{algorithmicx}
\usepackage{algorithm}
\usepackage{array}
\usepackage{fixltx2e}
\usepackage{stfloats}
\usepackage{url}
\usepackage{mathrsfs}
\usepackage[makeroom]{cancel}
\usepackage{lipsum}
\usepackage{graphicx}
\usepackage{bm}
\usepackage{amsbsy}
\usepackage{subfig}
\usepackage{algpseudocode} 
\usepackage[noadjust]{cite}

\makeatletter
\def\ps@IEEEtitlepagestyle{
	\def\@oddfoot{\mycopyrightnotice}
	\def\@evenfoot{}
}
\def\mycopyrightnotice{
	{\footnotesize
		\begin{minipage}{\textwidth}
			\centering
			Copyright~\copyright~2017 IEEE. Personal use of this material is permitted. However, permission to use this  \\ 
			material for any other purposes must be obtained from the IEEE by sending a request to pubs-permissions@ieee.org.
		\end{minipage}
	}
}
\begin{document}
%
\title{Adaptive Stimulus Selection in ERP-Based Brain-Computer Interfaces by Maximizing Expected Discrimination Gain}

\author{
	\IEEEauthorblockN{Dmitry Kalika\IEEEauthorrefmark{1}, Leslie M. Collins\IEEEauthorrefmark{1}\IEEEauthorrefmark{2}, Chandra S. Throckmorton\IEEEauthorrefmark{1}, Boyla O. Mainsah\IEEEauthorrefmark{1}}
	\IEEEauthorblockA{\IEEEauthorrefmark{1}Department of Electrical and Computer Engineering, Duke University, NC, USA 27708
		\\\IEEEauthorrefmark{2}Corresponding Author: Leslie.Collins@duke.edu}
}


%


\maketitle

\begin{abstract}
Brain-computer interfaces (BCIs) can provide  an alternative means of communication for individuals with severe neuromuscular limitations. The P300-based BCI speller relies on eliciting and detecting transient event-related potentials (ERPs) in electroencephalography (EEG) data, in response to a user attending to rarely occurring target stimuli amongst a series of non-target stimuli. However, in most P300 speller implementations, the stimuli to be presented are randomly selected from a limited set of options and stimulus selection and presentation are not optimized based on previous user data. In this work, we propose a data-driven method for stimulus selection based on the expected discrimination gain metric. The data-driven approach selects stimuli based on previously observed stimulus responses, with the aim of choosing a set of stimuli that will provide the most information about the user's intended target character. Our approach incorporates knowledge of physiological and system constraints imposed due to real-time BCI implementation. Simulations were performed to compare our stimulus selection approach to the row-column paradigm, the conventional stimulus selection method for P300 spellers. Results from the simulations demonstrated that our adaptive stimulus selection approach has the potential to significantly improve performance from the conventional method: up to 34\% improvement in accuracy and 43\% reduction in the mean number of stimulus presentations required to spell a character in a 72-character grid. In addition, our greedy approach to stimulus selection provides the flexibility to accommodate design constraints.
\end{abstract}


%
\IEEEpeerreviewmaketitle

\section{Introduction}\label{sec:intro}
Brain-computer interfaces (BCIs) can provide an alternative means of communication for individuals with severe neuromuscular limitations due to disease or physical trauma \cite{wolpaw_braincomputer_2002}.  One of the most widely researched non-invasive BCIs for communication is the P300-based BCI speller \cite{farwell_talking_1988,kubler_braincomputer_2009,mainsah_increasing_2015,sellers_p300-based_2006,sellers_brain-computer_2006}. The P300-based BCI relies on eliciting and detecting event-related potentials  (ERPs) embedded in electroencephalography (EEG) data in response to a user attending to a rarely occurring but relevant stimulus, termed target stimulus, amongst a series of irrelevant stimuli. The presentation of the rare target stimulus event elicits a specific ERP response, which is characterized by a large positive spike called the P300 signal \cite{sutton_evoked-potential_1965}. 

\begin{figure}
\includegraphics[scale=0.45]{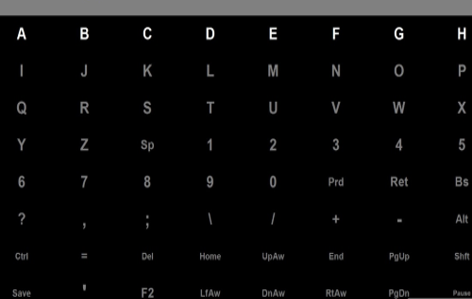}	 
\centering
\caption{P300 speller virtual keyboard with the first row of characters illuminated or flashed.}
\label{fig:p300VirtualGrid}
\end{figure}

In a visual P300 speller, the user is presented with a virtual keyboard, such as is shown in Figure \ref{fig:p300VirtualGrid}. To spell a desired character, the user focuses on that character, while subsets of characters, also called flash groups, are sequentially illuminated on the screen. Within this context, a flash group that contains the target character is termed a target stimulus, and ideally should elicit a P300 ERP when illuminated. An automated algorithm is used to analyze EEG data following the stimulus presentations to translate the user's responses into a BCI selection. However, the BCI decision-making process is error-prone due to the low signal-to-noise ratios (SNRs) of the elicited ERPs that are embedded in noisy EEG data. Typically, data from multiple stimulus presentations are analyzed to increase the ERP SNR for improved selection accuracy \cite{blankertz_single-trial_2011}. 

In most visual P300 spellers, the flash groups are selected at random from a set of limited options \cite{farwell_talking_1988,townsend_novel_2010}. In the conventional approach, known as the row-column (RC) paradigm, the flash groups are the rows and columns of characters arranged in a grid layout \cite{farwell_talking_1988}, such as shown in Figure \ref{fig:p300VirtualGrid}. Instead of randomly created flash groups, other methods have exploited error-correcting codes  from coding theory to optimize the composition and presentation order of flash groups, such as  in \cite{verhoeven_towards_2015,geuze_dense_2012,mainsah_performance-based_2017}. However, these approaches do not rely on previously observed EEG data to optimize stimulus selection. 

Only a few P300 speller studies have
implemented data-driven stimulus selection strategies. A data-driven approach selects stimuli based on previously observed stimulus responses; this allows the speller to select stimuli that can provide more information about the target character than random stimulus selection. Park et al. \cite{park_pomdp_2012} developed an adaptive approach to select row and column flash groups based on a partially observed Markov decision process (POMDP). Ma et al. \cite{ma_stochastic_2012} implemented a hierarchy of variable-sized flash groups based on a language model. In real-time BCI studies, both approaches resulted in significant performance improvements compared to the RC paradigm with random stimulus presentations. However, the POMPD approach becomes intractable for a real-time system when considering a search space considerably larger than row and column flash groups, and a hierarchical approach is easily susceptible to error propagation, especially when considering the amount of incorrect selections in users with low accuracy levels \cite{ma_stochastic_2012}. 

In this work, we propose an adaptive stimulus selection method based on the optimized sampling strategy developed by Kastella \cite{kastella_discrimination_1997}, which relies on the Kullback-Leibler divergence \cite{kullback_information_1951} or the discrimination gain function. We also propose a greedy approach to stimulus selection that considers an exponentially large search space to dynamically create flash groups, in order to provide the flexibility to accommodate constraints imposed for real-time BCI implementation.

The rest of this document is organized as follows. Section \ref{sec:Background} provides background on P300 speller operation, including our current BCI implementation. Section \ref{sec:adaptiveFlashing} outlines our proposed adaptive stimulus selection method. We present results from simulations in section \ref{sec:exps}, which demonstrate that our proposed approach has the potential to improve performance compared to the conventional RC paradigm.

\section{Background}\label{sec:Background}
In ERP-based BCIs, the goal of the BCI is to discern the character that the user intends to spell by distinguishing between target and non-target stimulus events. In the visual P300 speller, a flash group at time index $t$ can be represented as a binary vector, $\pmb{\mathscr{F}}_{t}\in [0,1]^M$, where $M$ is the number of possible BCI choices and the non-zero elements correspond to characters in the flash group. After stimulus presentation, a time window of EEG data is used to extract a feature vector, which is scored with a user-specific classifier to generate a score, $z_t$. Let $Z_t$ = $[z_1, z_2, \ldots , z_t]$ denote a sequence of classifier scores. The classifier scores are used to update a character scoring function that the BCI uses to evaluate how likely each of the BCI choices are the target character given the current data collection.

In this work, we use the naive Bayesian  dynamic stopping (DS) algorithm proposed in \cite{throckmorton_bayesian_2013}, where the scoring function is a probability distribution that is maintained over the character choices. Let $\textbf{P}_{t}\in [0,1]^M$  denote the vector of character probabilities at time index $t$. Following each stimulus presentation, $\pmb{\mathscr{F}}_{t}$, the resulting classifier score, $z_t$, is used to update $\textbf{P}_{t}$ accordingly: 

\begin{equation}\label{eq:BayesianUpdate}
\textbf{P}_{t,m} =\frac{\ell_{t,m}(z_t)\textbf{P}_{t-1,m}}{p(z_t|Z_{t-1})}= \frac{\ell_{t,m}(z_t)\textbf{P}_{t-1,m}}{\sum_{c=1}^M\ell_{t,c}(z_t)\textbf{P}_{t-1,c}}
\end{equation}
\begin{equation}\label{eq:likelihoodDists}
\ell_{t,m}(z_t)=
\begin{cases}
\ell0(z_t), & \pmb{\mathscr{F}}_{t,m}=0\\
\ell1(z_t), & \pmb{\mathscr{F}}_{t,m}=1
\end{cases}
\end{equation}
where $\textbf{P}_{t,m}$ and $\textbf{P}_{t-1,m}$ are the prior and posterior probabilities for character $C_m$; $\ell1$ and $\ell0$ are the target and non-target classifier score likelihood functions, respectively and $\ell_{t,m}(z_t)$ is the classifier score likelihood dependent on whether or not $C_m$ is present in the flash group, $ \pmb{\mathscr{F}}_{t}$. 

Data collection is stopped when the maximum character probability attains a pre-defined threshold value, $P_{th}$. For practical purposes, a data collection limit, $t_{max}$, is imposed since algorithm convergence within a reasonable amount of time is not always guaranteed. After data collection is stopped, the character with the maximum probability is selected as the user's intended target character.

During the Bayesian update process, new information from the classifier score is incorporated into the probability model, which updates the BCI system's belief of the user's target character. Our goal is to select and present flash groups that provide the most information to facilitate correct target character estimation. Kastella \cite{kastella_discrimination_1997} developed a data sampling strategy for improved multi-target detection and classification within a probabilistic framework. The approach is  based on maximizing the expected information gained with a hypothetical future sample, conditioned on all previously observed samples. In the next section, we describe how we exploit this sampling strategy for BCI  stimulus selection with the Bayesian dynamic stopping algorithm. 

\section{Proposed Adaptive Stimulus Selection Method}\label{sec:adaptiveFlashing}
In this section, we describe the data sampling strategy based on the expected discrimination gain function. The objective function (i.e., optimization goal) is presented in section \ref{sec:ObjectiveFunction} and the combinatorial optimization used to select a flash group is described in section \ref{sec:CombOpt}. Section \ref{sec:TTIOD} proposes modifications to the combinatorial optimization in order to incorporate system and physiological constraints.

\subsection{Objective Function}\label{sec:ObjectiveFunction}
Consider the Kullback-Leibler divergence \cite{kullback_information_1951} or the discrimination gain metric, which is a non-symmetric similarity measure between two probability distributions. For two discrete probability distributions, $\textbf{g}$ and $\textbf{q}$, the discrimination gain is: 

\begin{equation}\label{eq:KL}
KL(\textbf{g}||\textbf{q}) = \sum_{i=1}^{I}\textbf{g}_i\log{\left(\frac{\textbf{g}_i}{\textbf{q}_i}\right)}
\end{equation}

When using the naive Bayesian algorithm for character estimation (\ref{eq:BayesianUpdate}), the overall goal is to select a future flash group that maximizes the expected discrimination gain between the current probability distribution, $\textbf{P}_t$, and a hypothetical posterior distribution, $\textbf{P}^*_{t+1}$, over all possible future observations, conditioned on the current set of observations, $z^*_{t+1}|Z_t$.  We propose the following objective function for stimulus selection:
\begin{subequations}

\begin{align}
\Delta KL_t &= E_{z^*_{t+1}|Z_\textbf{t}}[KL(\textbf{P}_{t+1}(Z^*_{t+1})||\textbf{P}_t(Z_\textbf{t})]\\\label{eq:EKL2}
&= \int_{-\infty}^{\infty}\sum_{m=1}^{M}\textbf{P}^*_{t+1,m}\log{\left(\frac{\textbf{P}^*_{t+1,m}}{\textbf{P}_{t,m}}\right)}p(z^*_{t+1}|Z_t)dz^*_{t+1}
\end{align}
\begin{multline}\label{eq:EKL3}
= \int_{-\infty}^{\infty}\sum_{m=1}^{M}\ell_{t+1,m}(z^*_{t+1})\textbf{P}_{t,m}\\\log{\left(\frac{\ell_{t+1,m}(z^*_{t+1})}{\sum_{c=1}^M\ell_{t+1,c}(z^*_{t+1})\textbf{P}_{t,c}}\right)}dz^*_{t+1} 
\end{multline}
\end{subequations}
\\
\begin{equation}\label{eq:maxF}
\pmb{\mathscr{F}}^{\ddagger}_{t+1} = \underset{\pmb{\mathscr{F}}^*_{t+1}}{\arg\max} \Delta KL_t, \forall \pmb{\mathscr{F}}^*_{t+1} \in \pmb{\Omega}
\end{equation}
where $\Delta KL_t$ is the expected discrimination gain and $\pmb{\mathscr{F}}_{t+1}^{\ddagger}$ is the selected flash group that maximizes $\Delta KL_t$, given a search space of flash groups, $\pmb{\Omega}$.

The binary choice in the character likelihood assignments during  the Bayesian update process (\ref{eq:likelihoodDists})  allows us to simplify the evaluation of the integral in (\ref{eq:EKL2}) to (\ref{eq:EKL3}). The denominator in the log operator in  (\ref{eq:EKL3}) can be expressed as:

\begin{eqnarray}\label{eq:conditionalGrouped}
\sum_{c=1}^{M}\ell_{t+1,c}(z^*_{t+1})\textbf{P}_{t,c} =& \left[ \ell0(z^*_{t+1} )\sum_{\forall c:\pmb{\mathscr{F}}^*_{t+1,c}=0}\textbf{P}_{t,c}\right] \nonumber \\ +&\left[\ell1(z^*_{t+1} )\sum_{\forall c:\pmb{\mathscr{F}}^*_{t+1,c}=1}\textbf{P}_{t,c}\right]
\end{eqnarray}
where $\sum_{\forall c:\pmb{\mathscr{F}}^*_{t+1,c}=1}\textbf{P}_{t,c}$ is the sum of the prior probabilities at $t$ for the characters that are flashed at $t+1$. We  will define this summation as $P1_t$:
\begin{equation}\label{eq:p1}
P1_t = \sum_{\forall c:\pmb{\mathscr{F}}^*_{t+1,c}=1}\textbf{P}_{t,c}
\end{equation}
Similarly, we define $P0_t$ as:
\begin{equation}\label{eq:p0}
P0_t = \sum_{\forall c:\pmb{\mathscr{F}}^*_{t+1,c}=0}\textbf{P}_{t,c}
\end{equation}
Using (\ref{eq:p1}) and (\ref{eq:p0}) in (\ref{eq:conditionalGrouped}):
\begin{equation}\label{eq:groupedLikelihood2}
\sum_{c=1}^{M}\ell_{t+1,c}(z^*_{t+1})\textbf{P}_{t,c} = \ell0(z^*_{t+1})P0_t+\ell1(z^*_{t+1})P1_t
\end{equation}
From (\ref{eq:EKL3}) and (\ref{eq:groupedLikelihood2}), we obtain the following expression for the expected discrimination gain:
\begin{equation}\label{eq:KLSimple}
\Delta KL_t = \int_{-\infty}^{\infty}Sdz^*_{t+1}
\end{equation}

\begin{multline}\label{eq:S1}
S= \sum_{m=1}^{M}\ell_{t+1,m}(z^*_{t+1})\textbf{P}_{t,m}\\
 \log{\left(\frac{\ell_{t+1,m}(z^*_{t+1})}{\ell0(z^*_{t+1})P0_t+\ell1(z^*_{t+1})P1_t}\right)}
\end{multline}
where we introduce the integrand $S$ for notational simplicity. Similar to (\ref{eq:groupedLikelihood2}), we group flashed and non-flashed characters together and exploit the sum of probabilities, $P0_t = 1-P1_t$, to re-formularize $S$ as: 
\begin{flalign}\label{eq:SFinal}
S &= \nonumber\\ &P1_t\ell1(z^*_{t+1})\log\left(\frac{\ell1(z^*_{t+1})}{\ell0(z^*_{t+1})(1-P1_t)+\ell1(z^*_{t+1})P1_t}\right)\nonumber\\+
&(1-P1_t)\ell0(z^*_{t+1})\log\left(\frac{\ell0(z^*_{t+1})}{\ell0(z^*_{t+1})(1-P1_t)+\ell1(z^*_{t+1})P1_t}\right)
\end{flalign}

\begin{figure}
	\centering
	\includegraphics[scale=0.6]{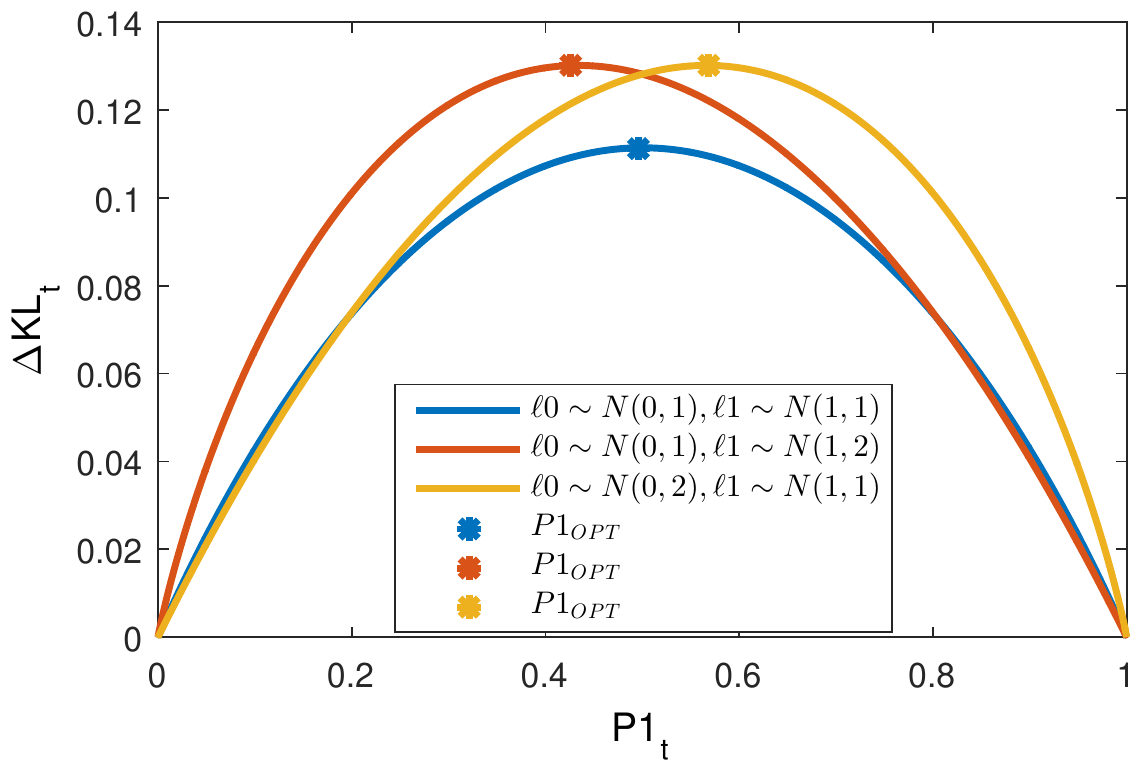}
	\caption{Three examples of pre-computed expected discrimination gain curves, $\Delta KL_t$ as a function of $P1_t$. $P1_{OPT}$ is the $P1_t$ that yields the maximum $\Delta KL_t$ for a given function.}
	\label{fig:expDiscCurve}
\end{figure}

We now have an expression for $\Delta KL_t$ based on the likelihood distributions $\ell0$ and $\ell1$, and the sum of probabilities, $P1_t$. It is important to note that when evaluating the integral in (\ref{eq:S1}) to determine $\Delta KL_t$, $z^*_{t+1}$ is being marginalized out over $\ell0$ and $\ell1$. Consequently, if $\ell0$ and $\ell1$ are defined, $\Delta KL_t$ depends only on $P1_t$. This simplification is particularly useful, as we do not always need to evaluate the integral in (\ref{eq:KLSimple}) to determine $\Delta KL_t$ for a proposed $\pmb{\mathscr{F}}^*_{t+1}$. Instead, we can pre-compute a $\Delta KL_t$ curve as a function of $P1_t$. Then, given a proposed $\pmb{\mathscr{F}}^*_{t+1}$, we determine $P1_t$ according to (\ref{eq:p1}) and then obtain $\Delta KL_t$ by indexing from the pre-computed curve. Some example $\Delta KL_t$ functions are shown in Figure \ref{fig:expDiscCurve}.

\subsection{Combinatorial Optimization}\label{sec:CombOpt}
Given a search space of flash groups, $\pmb{\Omega}$, we use (\ref{eq:maxF}) to determine the next flash group presentation, $\pmb{\mathscr{F}}_{t+1}^{\ddagger}$. For a fixed set of flash groups, such as row and column flash groups, $\pmb{\mathscr{F}}_{t+1}^{\ddagger}$ can be determined by selecting the flash group whose $P1_t$ maximizes $\Delta KL_t$. However, stimulus selection from a small search space may not always provide the most optimal or the best solution. We want the flexibility to dynamically create flash groups in order to better achieve our objective. However, given the exponentially large space of $2^M$ possibilities, an exhaustive search is impractical. 

To expand our search space, we will adopt a greedy approach for stimulus selection. We define the optimal $P1_t$ that maximizes $\Delta KL_t$ for specific $\ell0$ and $\ell1$ functions as $P_{OPT}$, as illustrated in Figure \ref{fig:expDiscCurve}. Given the current character probabilities, $\textbf{P}_t$, we iterate over an ordered list of probability values to construct flash groups that yield $P1_t$ values that are closest to $P_{OPT}$ from both sides. From the two choices, we select the flash group with maximum $\Delta KL_t$, $\widehat{\pmb{\mathscr{F}}}_{t+1}^{\ddagger}$.

\subsection{System and Physiological Constraints}\label{sec:TTIOD}
\begin{figure}
\centering
\includegraphics[scale=0.43]{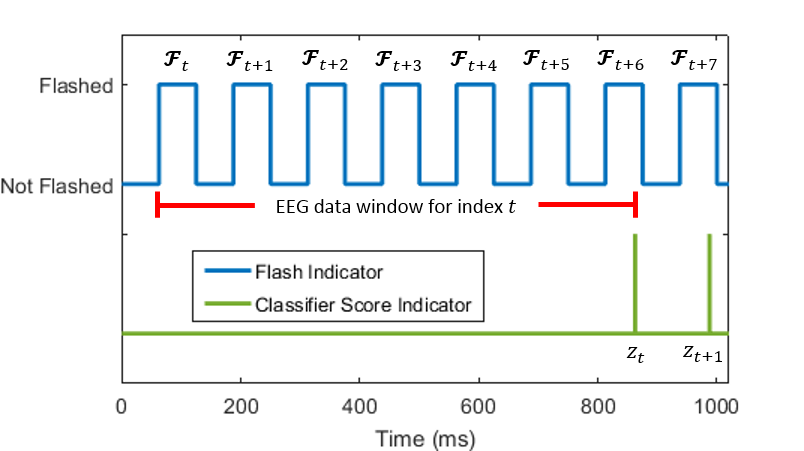}
\caption{Illustrative example to demonstrate the observation delay between a flash group's presentation, $\pmb{\mathscr{F}}_{t}$, and its resulting classifier score, $z_t$. }

\label{fig:obsDelayFig}
\end{figure}

In the previous section, it  was assumed that the classifier observation, $z_t$, was available prior to determining $\pmb{\mathscr{F}}_{t+1}^{\ddagger}$. However, this is typically not the case during online BCI implementation. Following each stimulus presentation, a time window of EEG data is analyzed to yield a classifier score. Consequently there is an \textit{observation delay} between the presentation of $\pmb{\mathscr{F}}_{t}$ and its associated classifier score, $z_t$, as illustrated in Figure \ref{fig:obsDelayFig}. Stimulus presentation is still ongoing during this delay. From Figure \ref{fig:obsDelayFig}, it can be observed that $\pmb{\mathscr{F}}_{t+6}$ will have already been presented prior to observing $z_t$. We will define the observation delay as OD = $\delta$, where  $\delta$ is the number of additional stimulus presentations prior to determining  $z_t$. In this work, we account for the observation delay by estimating $P_{t+\delta}$ recursively after observing $z_t$, and using $P_{t+\delta}$ to determine $\pmb{\mathscr{F}}_{t+\delta+1}^{\ddagger}$.

We also consider physiological limitations during stimulus selection. In particular, there are \textit{refractory effects} where the ERP SNR depends on the time interval between target stimulus presentations \cite{martens_overlap_2009}. Let the target-to-target interval (TTI) denote the number of stimulus events between target character presentations, such that a sequence TNNT denotes a TTI of 3, where N and T are non-target and target stimuli, respectively. It has been shown in the literature that short TTIs result in ERP responses with low SNR, and this negatively impacts target classification performance \cite{hill_effects_2009,verhoeven_towards_2015}. To alleviate refractory effects during stimulus selection, we will impose a minimum time interval $(TTI_{\min})$ between a character's presentation. 

\section{Numerical Experiment and Results}\label{sec:exps}
 Following the framework in \cite{mainsah2016using}, we perform numerical simulations of the P300 speller character selection process to compare various configurations of our proposed adaptive stimulus paradigm to the conventional RC paradigm. Assuming the speller grid shown in Figure \ref{fig:p300VirtualGrid}, during each iteration, the target character was chosen uniformly from 72 characters. Based on the flash groups defined by a stimulus paradigm condition, for non-target and target flash groups, the classifier scores were drawn according to normal distributions for $\ell0(z)$ and $\ell1(z)$, respectively, with parameters defined accordingly:
 
 \begin{equation}
 d^{\prime} = \frac{\mu_1-\mu_0}{\sigma}
 \end{equation} 
where $d^{\prime}$ is the detectability index as defined in \cite{birdsall_theory_1973}; $\mu_0$ and $\mu_1$ are the mean parameters for $\ell0(z)$ and $\ell1(z)$, respectively, and $\sigma$ is the common standard deviation. The Bayesian dynamic stopping algorithm (see section \ref{sec:Background}) was used for character selection, with uniform initialization probabilities, $P_{th} = 0.9$ and $t_{max}=120$ stimulus flashes. To provide a fair comparison with the RC paradigm condition, for the greedy adaptive paradigm, the maximum flash group size was set to 9 characters, equivalent to the size of a row flash group for the grid shown in Figure \ref{fig:p300VirtualGrid}. Selection accuracy and the average number of flashes prior to character selection, denoted as the expected stopping time (EST), are estimated as a function of $d^{\prime}$, with performance results averaged over 1500 iterations. 

In section \ref{sec:idealExp}, we present results from simulations with no constraints imposed. In section \ref{sec:realExp}, we present results from simulations where we consider realistic constraints imposed for online BCI implementation. 

\subsection{Simulations with Ideal Conditions}\label{sec:idealExp}
\begin{figure*}[!htb]
	\centering
	\subfloat[]{\includegraphics[width=2.5in]{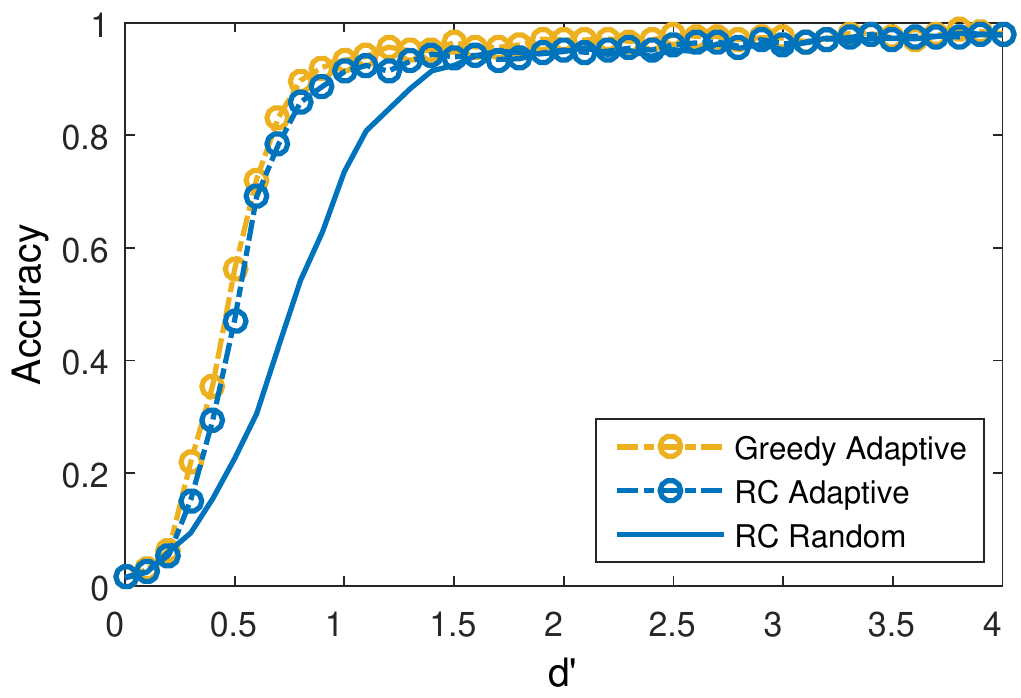}
		\label{fig:noRestrictAcc}}
	\hfil
	\subfloat[]{\includegraphics[width=2.5in]{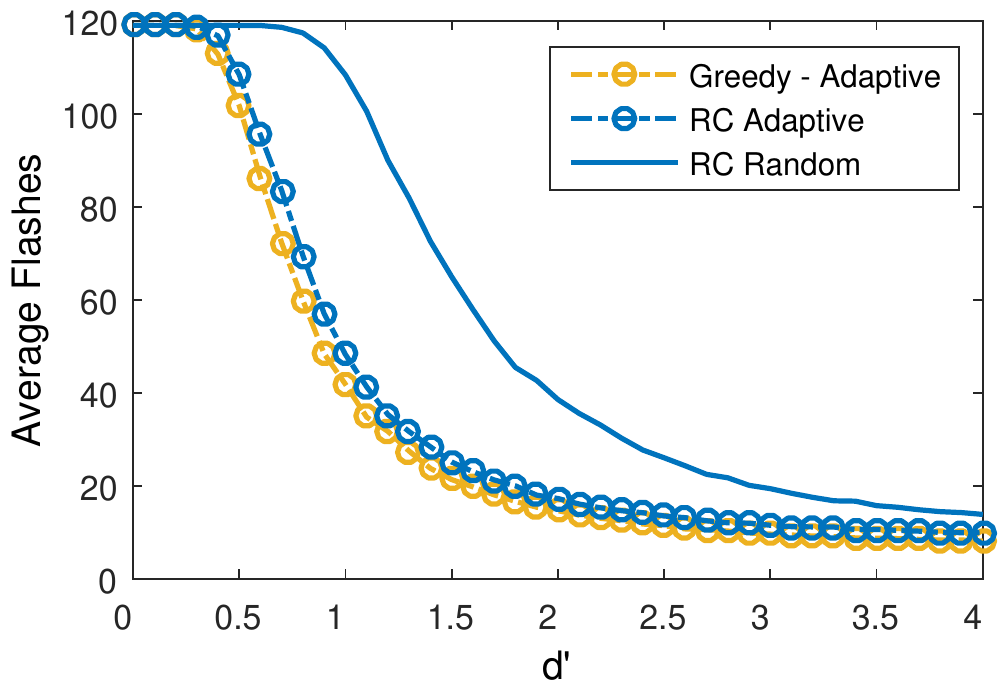}
		\label{fig:noRestrictFlashes}}
	\caption{Performance of the Bayesian dynamic stopping algorithm as a function of detectability index, $d^{\prime}$, for the RC random, and the RC adaptive and greedy adaptive paradigms, with no OD or TTI constraints: (a)  accuracy and (b) the expected stopping time, expressed in average number of flashes.}
	\label{fig:noRestrict}
\end{figure*}

 Assuming no OD or TTI constraints, we performed simulations to compare the performance of three stimulus presentation paradigms: RC random, RC adaptive, and greedy adaptive paradigms. In the RC random paradigm, row and column flash groups are randomly presented without replacement. In the RC adaptive paradigm, the search space was restricted to the row and column flash groups. We used the greedy algorithm described in section \ref{sec:CombOpt} for stimulus selection in the greedy adaptive paradigm. Figure \ref{fig:noRestrictAcc} and \ref{fig:noRestrictFlashes} show the accuracy and the EST, respectively. Substantial improvements in performance from the RC random paradigm are obtained with the adaptive stimulus paradigms, with increased accuracies of up to 41\% and decreased EST of up to 67\%. The performances of both adaptive stimulus paradigms are comparable; with the greedy adaptive paradigm performing slightly better. 

 The results of the adaptive paradigms with no constraints can be thought of as an upper performance bound for using an optimized stimulus selection strategy. In the next section, we examine the impact of imposing real-world constraints during the optimization process.
 
\subsection{Simulations with Realistic Online Conditions}\label{sec:realExp}
Adopting stimulus presentation parameters used in the online BCI study in \cite{mainsah_performance-based_2017}, we used observation delay OD = 6 and minimum target to target interval $TTI_{min} = 3$ as constraint parameters during stimulus selection. In a first set of simulations, we imposed only the observation delay when implementing the adaptive paradigms. In a second set of simulations, we also included the minimum TTI restriction. 
\begin{figure*}[!htb]
	\centering
	\subfloat[]{\includegraphics[width=2.5in]{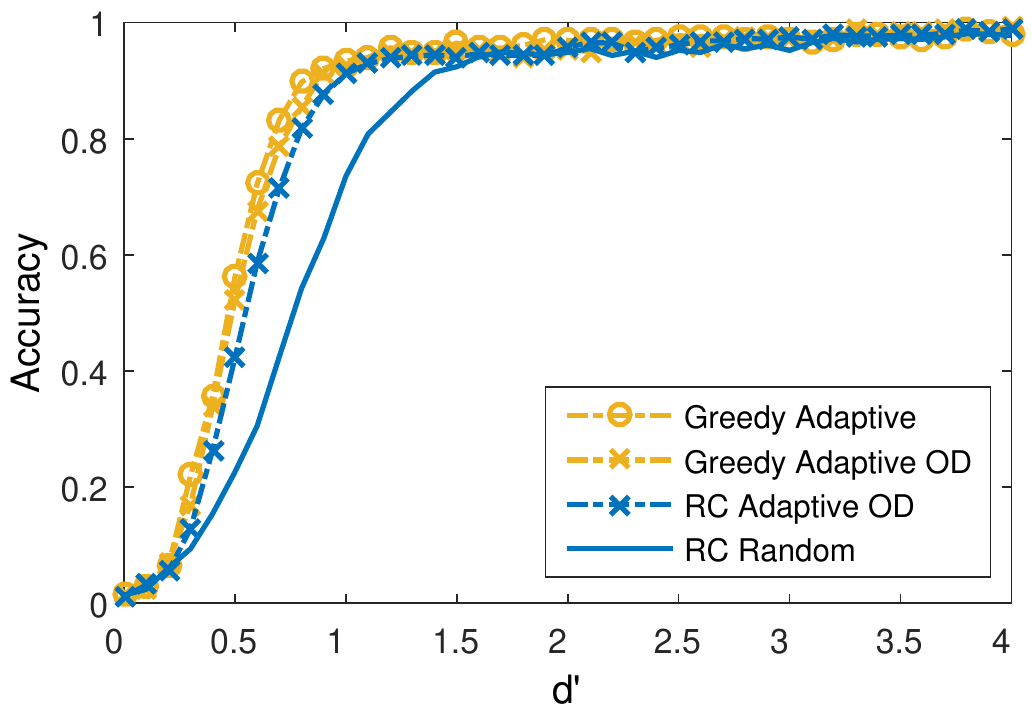}
		\label{fig:restrictODAcc}}
	\hfil
	\subfloat[]{\includegraphics[width=2.5in]{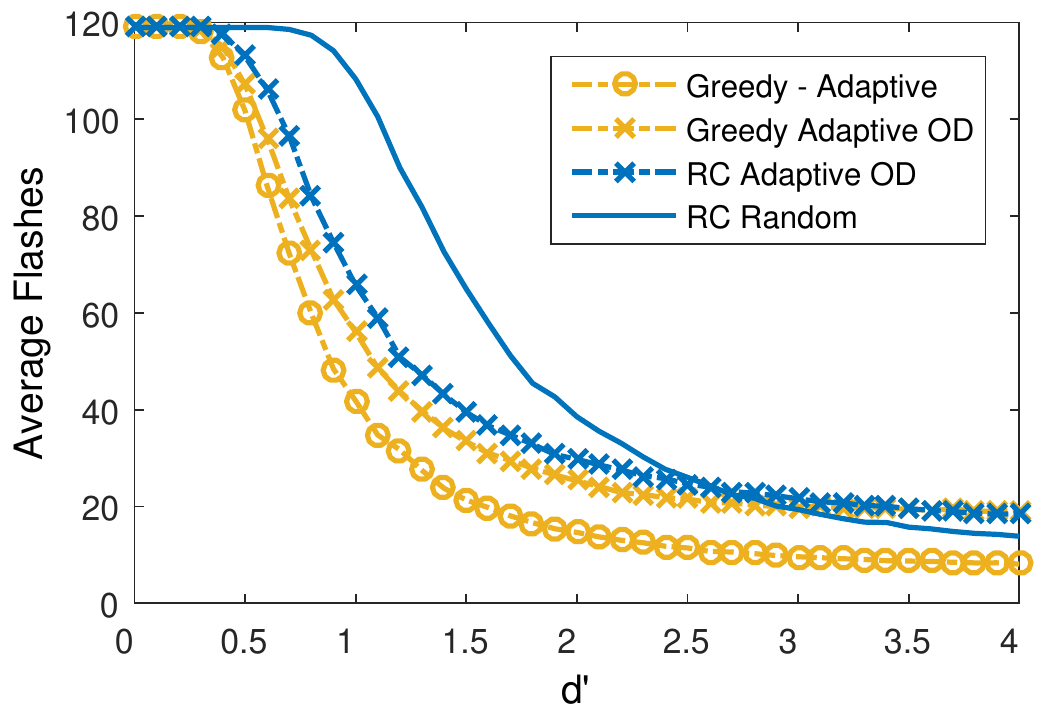}
		\label{fig:restrictODFlashes}}
	\caption{Performance of the Bayesian dynamic stopping algorithm as a function of detectability index, $d^{\prime}$, for the RC adaptive and greedy adaptive paradigms, with an observation delay of 6, (OD = 6): (a)  accuracy and (b) the expected stopping time, expressed in average number of flashes. The results for the RC random and greedy adaptive paradigms with no OD or TTI constraints are also shown.}
	\label{fig:restrictOD}
\end{figure*}
\begin{figure*}[!htb]
	\centering
	\subfloat[]{\includegraphics[width=2.5in]{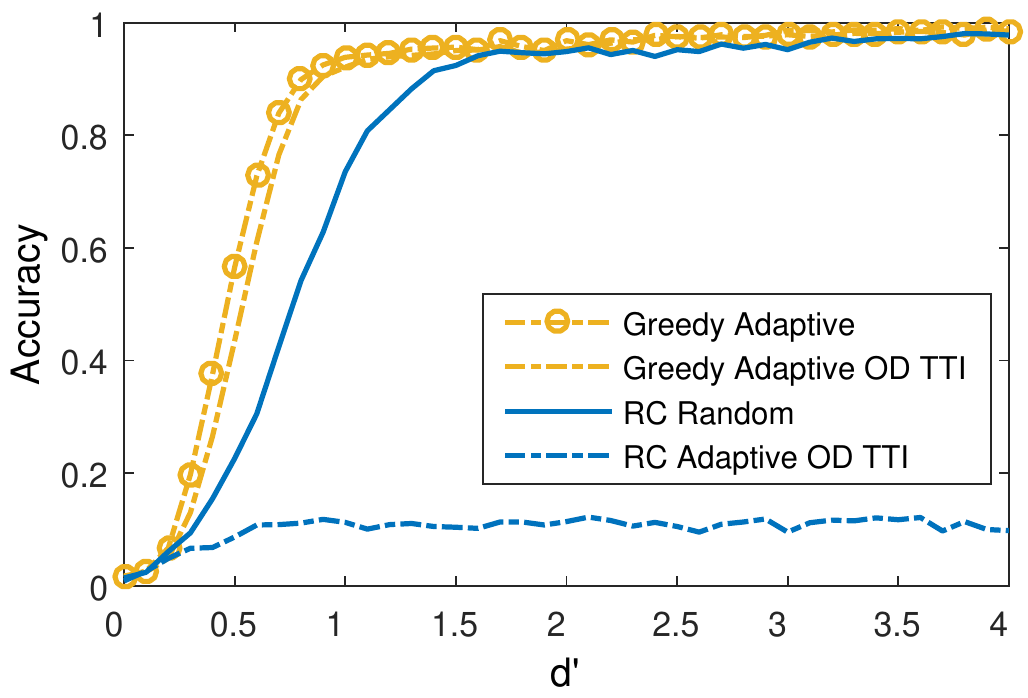}
		\label{fig:restrictODTTIAcc}}
	\hfil
	\subfloat[]{\includegraphics[width=2.5in]{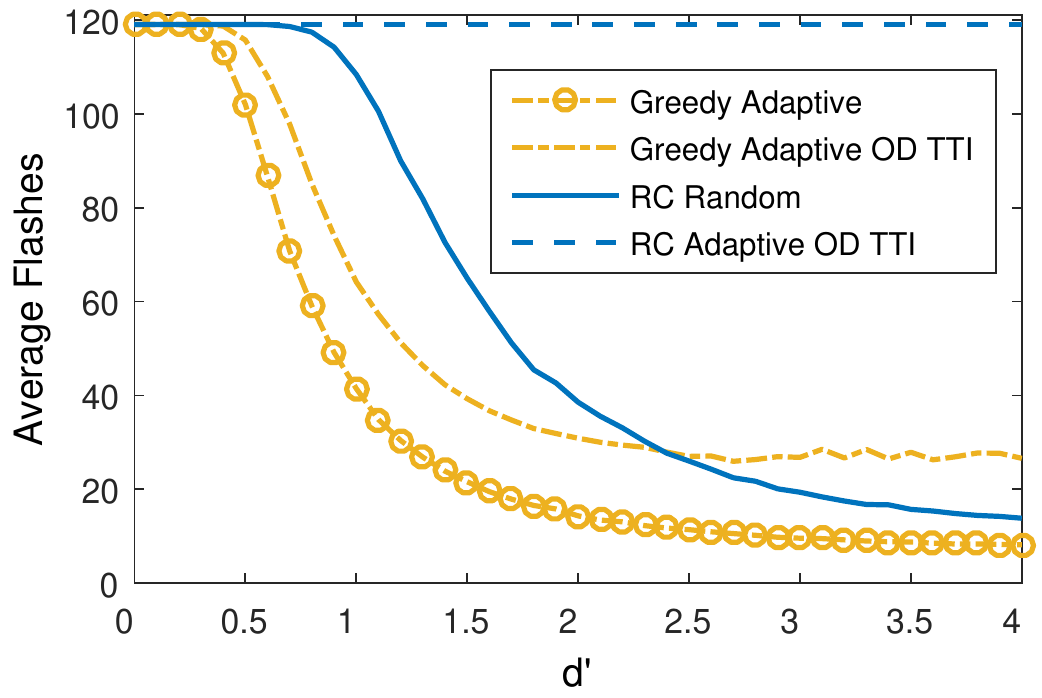}
		\label{fig:restrictODTTIFlashes}}
	\caption{Performance of the Bayesian dynamic stopping algorithm as a function of detectability index, $d^{\prime}$, for RC adaptive and greedy adaptive paradigms with an observation delay of 6 (OD = 6) and a minimum TTI restriction of 3 ($TTI_{min} = 3$): (a) accuracy and (b) the expected stopping time, expressed in average number of flashes. The results for the RC random and greedy adaptive paradigm with no OD or TTI constraints are also shown.}
	\label{fig:restrictTTIOD}
\end{figure*}
Figure \ref{fig:restrictOD} shows results when implementing the adaptive paradigms with an observation delay.  Compared to the previous case with no constraints, the adaptive paradigms with OD experience a drop in performance. The greedy approach provides a slight improvement over the RC adaptive paradigm. The adaptive paradigms with OD significantly outperform the RC random condition (accuracy increase of up to 37\% and a decrease in the EST of up to 51\% when $d^{\prime} <$ 3). It can also be observed that at $d^{\prime}>3$  where similar accuracy levels are observed across all paradigms, the EST with the adaptive paradigms with OD are slightly higher than the RC random paradigm. This demonstrates a potential negative impact on the spelling rates at high $d^{\prime}$ values when using a stimulus selection method that relies on delayed observations.

Figure \ref{fig:restrictTTIOD} shows results for the adaptive paradigms with the observation delay and a minimum TTI imposed. The RC adaptive paradigm with both constraints performs significantly worse than RC random despite utilizing all the available flashes, with an accuracy upper bound of $\approx$ 0.1. The greedy adaptive paradigm with OD and TTI constraints outperforms the RC random condition (accuracy increase of up to 34\% and a decrease in the EST of up to 43\% when $d^{\prime} <$ 2.5). While the accuracy of the greedy adaptive paradigm with both TTI and OD constraints is still comparable to the unconstrained case, the EST with the former is substantially increased compared to the latter. At $d^{\prime}>2.5$, the EST of the greedy adaptive paradigm with constraints is noticeably higher than the RC random condition. Nonetheless, the greedy approach is more robust to the imposition of constraints especially at low $d^{\prime}$ values. We are currently developing a method to decrease the EST at high $d^{\prime}$ values by imposing a specific TTI distribution for stimulus event presentations.

\section{Conclusions}
We have developed a data-driven adaptive method for stimulus selection in ERP-based BCIs based on maximizing the expected discrimination gain metric and provided an optimization function that relies on a single parameter. The data-driven approach allows the speller to select stimuli that can provide more information about the target character to improve BCI performance than random stimulus selection (e.g., row column paradigm). In addition, we proposed a greedy approach for stimulus selection when considering an exponentially large search space. Results from simulations demonstrated that significant performance improvements can potentially be obtained with the proposed adaptive stimulus selection method when compared to the conventional stimulus selection method. Furthermore, the flexibility of a greedy approach provided for more robustness when considering physiological and system constraints for real-time BCI implementation. 

In the future, we will implement the greedy adaptive paradigm on a real-time P300 speller system to validate our proposed approach using EEG data.

\section*{Acknowledgment}
This work was sponsored by NIH Grant \# R33 DC010470.



\bibliographystyle{IEEEtran}
\bibliography{IEEE_Cybernetics_2017_Conf}

\end{document}